\documentclass[aps,prl,showpacs,twocolumn,superscriptaddress,groupedaddress,floatfix,amsfonts]{revtex4}
\usepackage{graphicx}
\usepackage{dcolumn}
\usepackage{bm}
\usepackage{amssymb}
\usepackage{amsmath}
\usepackage{xspace}

\usepackage{color}
\newcommand{\CHECK}[1]{\textbf{\color{red}[#1]}\xspace}
\newcommand{\chk}[1]{\CHECK{CHECK THIS!}}

\def \qt  {$p_T^{\gamma\gamma}$}
\def \mass  {$M_{\gamma\gamma}$}
\def \dphi {$\Delta\phi_{\gamma\gamma}$}
\def \cost {$|\cos \theta^{*}|$}

\def \dqt  {$d\sigma/d$\mbox{\qt}}
\def \dmass  {$d\sigma/d$\mbox{\mass}}
\def \ddphi {$d\sigma/d$\mbox{\dphi}}
\def \dcost {$d\sigma/d$\mbox{\cost}}

\begin{document}


\hspace{5.2in} \mbox{FERMILAB-PUB-13-022-E}


\title{\boldmath Measurement of the differential cross sections for isolated direct photon pair production
in $p \bar p$ collisions at $\sqrt{s}=1.96$~TeV}


%
\affiliation{LAFEX, Centro Brasileiro de Pesquisas F\'{i}sicas, Rio de Janeiro, Brazil}
\affiliation{Universidade do Estado do Rio de Janeiro, Rio de Janeiro, Brazil}
\affiliation{Universidade Federal do ABC, Santo Andr\'e, Brazil}
\affiliation{University of Science and Technology of China, Hefei, People's Republic of China}
\affiliation{Universidad de los Andes, Bogot\'a, Colombia}
\affiliation{Charles University, Faculty of Mathematics and Physics, Center for Particle Physics, Prague, Czech Republic}
\affiliation{Czech Technical University in Prague, Prague, Czech Republic}
\affiliation{Center for Particle Physics, Institute of Physics, Academy of Sciences of the Czech Republic, Prague, Czech Republic}
\affiliation{Universidad San Francisco de Quito, Quito, Ecuador}
\affiliation{LPC, Universit\'e Blaise Pascal, CNRS/IN2P3, Clermont, France}
\affiliation{LPSC, Universit\'e Joseph Fourier Grenoble 1, CNRS/IN2P3, Institut National Polytechnique de Grenoble, Grenoble, France}
\affiliation{CPPM, Aix-Marseille Universit\'e, CNRS/IN2P3, Marseille, France}
\affiliation{LAL, Universit\'e Paris-Sud, CNRS/IN2P3, Orsay, France}
\affiliation{LPNHE, Universit\'es Paris VI and VII, CNRS/IN2P3, Paris, France}
\affiliation{CEA, Irfu, SPP, Saclay, France}
\affiliation{IPHC, Universit\'e de Strasbourg, CNRS/IN2P3, Strasbourg, France}
\affiliation{IPNL, Universit\'e Lyon 1, CNRS/IN2P3, Villeurbanne, France and Universit\'e de Lyon, Lyon, France}
\affiliation{III. Physikalisches Institut A, RWTH Aachen University, Aachen, Germany}
\affiliation{Physikalisches Institut, Universit\"at Freiburg, Freiburg, Germany}
\affiliation{II. Physikalisches Institut, Georg-August-Universit\"at G\"ottingen, G\"ottingen, Germany}
\affiliation{Institut f\"ur Physik, Universit\"at Mainz, Mainz, Germany}
\affiliation{Ludwig-Maximilians-Universit\"at M\"unchen, M\"unchen, Germany}
\affiliation{Fachbereich Physik, Bergische Universit\"at Wuppertal, Wuppertal, Germany}
\affiliation{Panjab University, Chandigarh, India}
\affiliation{Delhi University, Delhi, India}
\affiliation{Tata Institute of Fundamental Research, Mumbai, India}
\affiliation{University College Dublin, Dublin, Ireland}
\affiliation{Korea Detector Laboratory, Korea University, Seoul, Korea}
\affiliation{CINVESTAV, Mexico City, Mexico}
\affiliation{Nikhef, Science Park, Amsterdam, the Netherlands}
\affiliation{Radboud University Nijmegen, Nijmegen, the Netherlands}
\affiliation{Joint Institute for Nuclear Research, Dubna, Russia}
\affiliation{Institute for Theoretical and Experimental Physics, Moscow, Russia}
\affiliation{Moscow State University, Moscow, Russia}
\affiliation{Institute for High Energy Physics, Protvino, Russia}
\affiliation{Petersburg Nuclear Physics Institute, St. Petersburg, Russia}
\affiliation{Instituci\'{o} Catalana de Recerca i Estudis Avan\c{c}ats (ICREA) and Institut de F\'{i}sica d'Altes Energies (IFAE), Barcelona, Spain}
\affiliation{Uppsala University, Uppsala, Sweden}
\affiliation{Taras Shevchenko National University of Kyiv, Kiev, Ukraine}
\affiliation{Lancaster University, Lancaster LA1 4YB, United Kingdom}
\affiliation{Imperial College London, London SW7 2AZ, United Kingdom}
\affiliation{The University of Manchester, Manchester M13 9PL, United Kingdom}
\affiliation{University of Arizona, Tucson, Arizona 85721, USA}
\affiliation{University of California Riverside, Riverside, California 92521, USA}
\affiliation{Florida State University, Tallahassee, Florida 32306, USA}
\affiliation{Fermi National Accelerator Laboratory, Batavia, Illinois 60510, USA}
\affiliation{University of Illinois at Chicago, Chicago, Illinois 60607, USA}
\affiliation{Northern Illinois University, DeKalb, Illinois 60115, USA}
\affiliation{Northwestern University, Evanston, Illinois 60208, USA}
\affiliation{Indiana University, Bloomington, Indiana 47405, USA}
\affiliation{Purdue University Calumet, Hammond, Indiana 46323, USA}
\affiliation{University of Notre Dame, Notre Dame, Indiana 46556, USA}
\affiliation{Iowa State University, Ames, Iowa 50011, USA}
\affiliation{University of Kansas, Lawrence, Kansas 66045, USA}
\affiliation{Louisiana Tech University, Ruston, Louisiana 71272, USA}
\affiliation{Northeastern University, Boston, Massachusetts 02115, USA}
\affiliation{University of Michigan, Ann Arbor, Michigan 48109, USA}
\affiliation{Michigan State University, East Lansing, Michigan 48824, USA}
\affiliation{University of Mississippi, University, Mississippi 38677, USA}
\affiliation{University of Nebraska, Lincoln, Nebraska 68588, USA}
\affiliation{Rutgers University, Piscataway, New Jersey 08855, USA}
\affiliation{Princeton University, Princeton, New Jersey 08544, USA}
\affiliation{State University of New York, Buffalo, New York 14260, USA}
\affiliation{University of Rochester, Rochester, New York 14627, USA}
\affiliation{State University of New York, Stony Brook, New York 11794, USA}
\affiliation{Brookhaven National Laboratory, Upton, New York 11973, USA}
\affiliation{Langston University, Langston, Oklahoma 73050, USA}
\affiliation{University of Oklahoma, Norman, Oklahoma 73019, USA}
\affiliation{Oklahoma State University, Stillwater, Oklahoma 74078, USA}
\affiliation{Brown University, Providence, Rhode Island 02912, USA}
\affiliation{University of Texas, Arlington, Texas 76019, USA}
\affiliation{Southern Methodist University, Dallas, Texas 75275, USA}
\affiliation{Rice University, Houston, Texas 77005, USA}
\affiliation{University of Virginia, Charlottesville, Virginia 22904, USA}
\affiliation{University of Washington, Seattle, Washington 98195, USA}
\author{V.M.~Abazov} \affiliation{Joint Institute for Nuclear Research, Dubna, Russia}
\author{B.~Abbott} \affiliation{University of Oklahoma, Norman, Oklahoma 73019, USA}
\author{B.S.~Acharya} \affiliation{Tata Institute of Fundamental Research, Mumbai, India}
\author{M.~Adams} \affiliation{University of Illinois at Chicago, Chicago, Illinois 60607, USA}
\author{T.~Adams} \affiliation{Florida State University, Tallahassee, Florida 32306, USA}
\author{G.D.~Alexeev} \affiliation{Joint Institute for Nuclear Research, Dubna, Russia}
\author{G.~Alkhazov} \affiliation{Petersburg Nuclear Physics Institute, St. Petersburg, Russia}
\author{A.~Alton$^{a}$} \affiliation{University of Michigan, Ann Arbor, Michigan 48109, USA}
\author{V.B.~Anikeev} \affiliation{Institute for High Energy Physics, Protvino, Russia}
\author{A.~Askew} \affiliation{Florida State University, Tallahassee, Florida 32306, USA}
% ...
%
%
\collaboration{The D0 Collaboration\footnote{with visitors from
$^{a}$Augustana College, Sioux Falls, SD, USA,
$^{b}$The University of Liverpool, Liverpool, UK,
$^{c}$UPIITA-IPN, Mexico City, Mexico,
$^{d}$DESY, Hamburg, Germany,
$^{e}$SLAC, Menlo Park, CA, USA,
$^{f}$University College London, London, UK,
$^{g}$Centro de Investigacion en Computacion - IPN, Mexico City, Mexico,
$^{h}$ECFM, Universidad Autonoma de Sinaloa, Culiac\'an, Mexico,
$^{i}$Universidade Estadual Paulista, S\~ao Paulo, Brazil,
$^{j}$Karlsruher Institut f\"ur Technologie (KIT) - Steinbuch Centre for Computing (SCC)
and
$^{k}$Office of Science, U.S. Department of Energy, Washington, D.C. 20585, USA.
}} \noaffiliation
\vskip 0.25cm
\date{\today}

\begin{abstract}
We present measurements of direct photon pair production cross sections using 8.5 fb$^{-1}$ of data collected with the D0 detector 
at the Fermilab Tevatron $p \bar p$ collider. The results are presented as differential distributions of 
the photon pair invariant mass $d\sigma/dM_{\gamma \gamma}$, pair transverse momentum $d \sigma /dp^{\gamma \gamma}_T$, 
azimuthal angle between the photons $d\sigma/d\Delta \phi_{\gamma \gamma}$, and polar scattering angle 
in the Collins-Soper frame $d\sigma /d|\cos \theta^*|$. Measurements are performed 
for isolated photons with transverse momenta $p^{\gamma}_T>18 ~(17)$ GeV for the leading (next-to-leading) photon in $p_T$,
pseudorapidities $|\eta^{\gamma}|<0.9$, and a separation in $\eta-\phi$ space $\Delta\mathcal R_{\gamma\gamma} > 0.4$.
We present comparisons with the predictions from Monte Carlo event generators {\sc diphox} and {\sc resbos}
implementing QCD calculations at next-to-leading order, $2\gamma${\sc nnlo} at
next-to-next-to-leading order, and {\sc sherpa} using matrix elements with higher-order real emissions matched to parton shower.
\end{abstract}

\pacs{13.85.Qk, 12.38.Qk}
\maketitle



Precise knowledge of the direct diphoton (DDP) production differential cross section is
a cornerstone of the search for the standard model (SM) Higgs boson by experiments at the Large Hadron Collider~\cite{atlas_higgs, cms_higgs} 
and the Tevatron~\cite{Tev_higgs, cdf_higgs}.
The term ``direct'' means that these photons do not result from meson, for example, 
$\pi^0,\eta,\omega$, or $K^0_{\rm S}$ decays. 
DDP production is also a significant background in searches for Kaluza-Klein~\cite{d0_led} or 
Randall-Sundrum~\cite{cdf_led} gravitons decaying into two photons, as well as other 
new phenomena processes, such as decays of heavy resonances~\cite{mrenna} 
or cascade decays of supersymmetric particles~\cite{supersymmetry}.
For these searches, DDP production is an irreducible background, and it is crucial to have a detailed understanding
of the distributions of key kinematic variables \cite{resbos}.  

In addition to investigating physics beyond the SM,
DDP production processes are important for studying quantum chromodynamics (QCD) and
measuring parton distribution functions (PDFs).
DDP production cross sections have been examined at fixed-target \cite{fix1, fix2} and 
collider experiments \cite{ua1, ua2, d0_dpp, cdf_dpp, atlas_dpp, cms_dpp}. 
%
DDP events at the Tevatron $p \bar p$ collider are produced predominantly
through quark-antiquark annihilation $q \bar q \rightarrow \gamma \gamma$ 
and gluon-gluon fusion ($gg \rightarrow \gamma \gamma$) via a quark-loop diagram.
The  matrix element (ME) for the latter process is suppressed by $\alpha_s^2$ relative
to $q \bar q$ annihilation, but 
its total production rate at low $\gamma\gamma$ invariant mass (\mass) 
and intermediate $\gamma\gamma$ transverse momentum (\qt) is quite 
significant due to the relatively large values of the gluon PDFs in that kinematic region.
By the same argument,  gluon-gluon fusion 
becomes even more important at the LHC~\cite{diphox}.
DDP events may also originate from processes such as
$qg \rightarrow q \gamma, q\bar q \to g\gamma$,  and $gg \rightarrow q \bar q$, 
where a photon with large transverse momentum is radiated from the final state parton.
These processes, being nearly collinear, require the introduction of a fragmentation function in 
perturbative QCD (pQCD) calculations~\cite{diphox}. 
Photon isolation requirements reduce the contribution of such fragmentation events.
However, their contribution may be still quite large at 
low  $\gamma \gamma$ azimuthal angle difference (\dphi) and 
for intermediate \qt ~\cite{diphox,diphox_tev}, which is the DDP transverse momentum. 



In this Letter, we present measurements of differential cross sections of DDP production 
using the dataset collected at the Fermilab Tevatron D0 experiment between June 2006 and September 2011.
The dataset corresponds to an integrated luminosity of $8.5 \pm 0.5$ fb$^{-1}$~\cite{d0lumi}.

Measurements are performed 
%
as functions of \mass, \qt, \dphi, and \cost, the absolute value of the cosine of the polar scattering 
angle of the diphoton system in the Collins-Soper frame~\cite{collins}. 
Here we approximate \cost ~by $|\tanh[(\eta_1-\eta_2)/2]|$, where $\eta_{1, 2}$ are the pseudo-rapidities~\cite{d0_coordinate} of the leading 
and next-to-leading photons ranked by $p_T$.  These four
variables emphasize different phenomena in the diphoton production mechanism.
\mass \space usually serves as a probe for new phenomena searches~\cite{atlas_higgs, cms_higgs, d0_led, cdf_led, mrenna} and PDFs. 
The \qt \space and \dphi \space shapes are mostly sensitive to the initial state gluon radiation and fragmentation effects. 
The \cost \space angle is sensitive to PDFs and spin correlations in the final state. 
In contrast with the previous D0 measurement \cite{d0_dpp}, in this analysis we do not impose explicit minimum
requirements on \mass ~or \dphi, ~nor do we require that \mass$>p_{T}^{\gamma\gamma}$, making the measurements more universal.
By separating the data into two subsets, with \dphi$\geq\pi/2$ and \dphi$<\pi/2$, 
we isolate regions with  smaller and larger expected relative contributions
from the fragmentation processes.

We compare our results with the theoretical predictions generated using the  {\sc diphox} \cite{diphox}, 
{\sc resbos} \cite{resbos,resbos2,resbos3},
$2\gamma${\sc nnlo} \cite{2gnnlo} and {\sc sherpa} \cite{sherpa} event generators. 
The general multipurpose generator approach is to employ interleaved QCD and 
quantum electrodynamics (QED) parton shower (PS) to describe initial and final state radiation.
The {\sc sherpa} Monte Carlo (MC) event generator improves this technique by
including higher-order real-emission matrix elements~\cite{sherpa_gam}. 
Matching between partons coming from real emissions in the ME and jets from PS
is done at some (hardness) scale $Q_{\rm cut}$ defined following the prescriptions given in Ref.~\cite{sherpa_gam}.
We use events generated with all MEs with two photons and up to two hard partons.
However, the ME for gluon-gluon scattering $gg \rightarrow \gamma \gamma$ in {\sc sherpa} does not have
real parton emissions.
As is shown in Ref.~\cite{sherpa_gam}, {\sc sherpa} provides a good description of the fragmentation function measured
at LEP at high fraction of the jet energy carried by the photon, corresponding to tight photon isolation cuts.
The loop corrections matching the higher order MEs are missing in {\sc sherpa}, 
which can make predictions significantly scale-dependent and 
may lead to underestimation of $\gamma \gamma$ rates.
In the {\sc sherpa} version used in this paper~\cite{sherpa}, the inherent
next-to-leading-logarithmic effect of correlated emissions is invoked
in parton-shower simulations by appropriately choosing a scale factor
for the argument of the running strong coupling constant~\cite{CWM, BSZ, PC}.
%
%
The {\sc diphox} and {\sc resbos} packages 
provide predictions at next-to-leading order (NLO) in pQCD, 
with the $gg\to \gamma\gamma$ process considered only at the leading order approximation in {\sc diphox}. 
Also, in {\sc diphox},  
explicit single and double parton-to-photon fragmentation processes are included at NLO accuracy,
while in {\sc resbos}, rates of fragmentation processes are approximated by a function.
Only in {\sc resbos} are the effects of soft and collinear
initial state gluon emissions resummed to all orders \cite{resbos3}. 
The resummation should be important for a correct description of the $p_T^{\gamma\gamma}$ distribution close to zero
and the $\Delta\phi_{\gamma\gamma}$ distribution close to $\pi$.
%
The $2\gamma${\sc nnlo} generator, which appeared recently, exploits the \qt \space subtraction formalism \cite{qtSubtr} 
that handles the unphysical infra-red divergences up to next-to-next-to-leading order (NNLO).
It takes into account most diagrams ($q\bar{q}$ and $qg$ scatterings) at $\mathcal O(\alpha_s^2)$ accuracy; 
however, in the current calculations, there is no higher-order correction to the $gg\to \gamma\gamma$ box diagram 
and no soft gluon resummation is  applied.
Additionally, it does not take into account the fragmentation contributions. 

The D0 detector, where the DDP measurements are performed, is a general purpose detector described in detail elsewhere~\cite{d0det, l1cal, l0}. 
The sub-detectors used in this analysis to trigger events and reconstruct photons are the calorimeter, 
the central tracking system, and the central preshower. The muon detection system is used to compare 
data and MC simulation sets of $Z \rightarrow \mu^+ \mu^- + \gamma$ events 
to obtain data-to-MC scale factors for reconstruction efficiency. The central tracking system, used to reconstruct tracks of 
charged particles, consists of a silicon micro-strip detector (SMT) and a central fiber track detector (CFT), 
both embedded in a 2~T solenoidal magnetic field. The solenoid is surrounded by the central preshower (CPS) 
detector located immediately before the inner layer of the electromagnetic calorimeter. The CPS consists of 
approximately one radiation length of lead absorber surrounded by three layers of scintillating strips. 
The calorimeter is composed of three sections: a central section covering the range of pseudo-rapidities 
$|\eta_{\rm det}|<1.1$~\cite{d0_coordinate} and two end calorimeters (EC) with coverage extending to $|\eta_{\rm det}|\approx4.2$,
with all three housed in separate cryostats.
The electromagnetic (EM) calorimeter is composed of four layers of 
$\Delta \eta_{\rm det} \times \Delta \phi_{\rm det}=0.1 \times 0.1$ cells, with the exception of layer three (EM3) 
with $0.05 \times 0.05$ granularity. The calorimeter resolution for measurements
 of the electron/photon energy at 50 GeV is about 3.6\%. The energy response of the calorimeter to photons is calibrated 
using electrons from $Z$ boson decays. Since electrons and photons shower differently in matter, additional 
corrections as a function of $\eta$ are derived using a detailed {\sc geant}-based~\cite{geant} simulation of 
the D0 detector response. These corrections are the largest, (2.0--2.5)\%, at low photon energies ($\approx 20$ GeV). 
Events satisfying the following trigger requirements are recorded: at least two clusters of energy in the EM calorimeter
with a loose shower shape requirement and a range of $p_T$ thresholds between 15 GeV and 25 GeV. 
Luminosity is measured using plastic scintillator arrays placed in front of the EC cryostats.

Events are selected with at least two photon candidates with transverse momentum $p_T>18 ~(17)$ GeV for the leading (next-to-leading) 
candidate and pseudorapidity $|\eta|<0.9$.
We require a slight difference between the $p_T$ cutoffs for the two photons to avoid a divergent 
kinematic region of the NLO calculations \cite{diphox}. The trigger is more than $90\%$ efficient for these selections. 

%
At high instantaneous luminosities there is more than one $p\bar p$ interaction 
per beam crossing. The photon $p_T$ is computed with respect to the reconstructed 
$p\bar p$ interaction vertex with the highest number of associated tracks, 
called the event vertex \cite{d0_dpp}. The event vertex is required to be 
reconstructed within 60 cm of the center of the detector along the beam axis ($z$), 
and satisfies this requirement in 98\% of events. 
We correct for effects of selecting an incorrect vertex (in about 35\% of events) 
using $Z \rightarrow e^+ e^-$ data events, where we remove tracks corresponding 
to the electron and positron to model DPP production.


Photon candidates are formed from calorimeter towers in a cone of radius $\mathcal R = \sqrt{(\Delta \eta)^2+(\Delta \phi)^2}=0.4$ 
around a seed tower~\cite{d0det}. A stable cone is found iteratively, 
and the final cluster energy is recalculated from the inner core 
within $\mathcal R = 0.2$. The photon candidates are required to: (i) have $\geq 97\%$ of the cluster energy 
deposited in the EM calorimeter layers; (ii) be isolated in the calorimeter according to $[E_{\rm tot}(0.4)-E_{\rm EM}(0.2)]/E_{\rm EM}(0.2)<0.07$, 
where $E_{\rm tot}(\mathcal R)$ $[E_{\rm EM}(\mathcal R)]$ is the total [EM only] energy in a cone of radius $\mathcal R$; 
(iii) have the scalar sum of $p_T$'s of all tracks originating from the event vertex in an annulus of $0.05<\mathcal{R}<0.4$ 
around the EM cluster less than $1.5$ GeV; 
and (iv) have an energy-weighted EM shower width consistent with that expected for an electromagnetic shower. 
To suppress electron misidentification as photons, the EM clusters are required to have no spatial match 
to a charged particle track or any tracker hit configuration consistent with an electron. 
The two photon EM clusters are required to be separated by $\Delta \mathcal R_{\gamma\gamma}>0.4$.
 
Additional group of variables exploiting the differences between the photon-initiated and jet activity in the EM calorimeter 
and the tracker are combined into an artificial neural network (NN) to further reject jet background~\cite{nn}. 
In these background events, photons are mainly produced from decays of energetic $\pi^0$ and $\eta$ mesons.
The NN is trained on $\gamma$ and jet {\sc pythia}~\cite{pythia} MC samples. 
The generated MC events are processed through 
a {\sc geant}-based simulation of the D0 detector. Simulated events are overlaid with data events from random $p\bar{p}$ crossings 
to properly model the effects of multiple $p \bar p$ interactions and detector noise in data. 
Care is taken to ensure that the luminosity distribution 
in the overlay events is similar to the data used in the analysis. 
MC events are then processed through the same reconstruction procedure as the data. 
MC events are reweighted to take into account the trigger efficiency in data, and
small observed differences in instantaneous luminosity and distribution of the $z$ coordinate of the event vertex.
Photon radiation from charged leptons 
in $Z$ boson decays ($Z \rightarrow \ell^+ \ell^- \gamma, \mbox{\space}\ell=e,\mu$) is used to validate the NN performance~\cite{nnvalid}. 
The NN describes the data well and gives significant extra discrimination against jets. 
The photon candidates in this analysis are chosen such that their NN output requirement retains 98\%  
of photons and rejects $\approx 40\%$ of jets beyond the rejection provided by the selection described above~\cite{d0_dpp}.


We estimate contributions from instrumental $\gamma+$jet and dijet backgrounds and also the contribution 
from $Z$ boson/Drell-Yan production events $Z/\gamma^* \rightarrow e^+ e^-$ (ZDY). In the instrumental backgrounds, 
one or more jets are misidentified as photons from jet-forming partons that hadronize into isolated neutral meson(s)
($\pi^0$ or $\eta$) giving rise to two or more photons in the final state. 
Electrons in the ZDY background 
can be misidentified as photons due to similarities in the shower shape.
The contribution from the ZDY events is estimated from MC simulation with {\sc pythia}, normalized to the NNLO cross section~\cite{zdynnlo}. 
On average, 2\% of the electrons survive the selection criteria above, mainly due to the inefficiency of matching a charged track 
to an electron. In data this inefficiency is higher than in MC and the ZDY contribution is corrected for these differences.

The $\gamma+$jet and dijet instrumental backgrounds are estimated by fitting a two-dimensional (2D) distribution of the leading 
and next-to-leading photon NN outputs  with templates extracted from DDP {\sc sherpa} signal
and EM-jet {\sc pythia} MC samples. 
In the latter, constraints are placed at the generator level to increase the statistics of jet events fluctuating 
into EM-like objects~\cite{nn}. For the $\gamma+$jet template, the photon candidate is taken from either 
the $\gamma \gamma$ sample or from the EM-jet sample, while for the dijet template, both candidates are taken from the EM-jet sample. 
Table~\ref{databkg} shows the numbers of events surviving the selection in data for different \dphi ~regions, 
as well as the number of data events from each of the four sources as determined by a fit of the signal and
background templates to data. The typical DDP purity in the selected data events is around 60\%.

\begin{table}[htb]
  \centering
  \caption{\label{databkg} \small
    The numbers of $\gamma\gamma$ ($N_{\gamma \gamma}$), $\gamma j+ j\gamma$ ($N_{\gamma j}$), 
    $jj$ ($N_{jj}$), and ZDY ($N_{\rm ZDY}$) events and their total. 
    The quoted uncertainties are statistical only and 
    for $N_{\gamma \gamma}$, $N_{\gamma j}$, and $N_{jj}$ are from 2D fitting. 
    }
  \begin{tabular}{cccc}
    \hline 
                  & Full \dphi & \dphi$<\pi/2$ & \dphi$\geq\pi/2$ \\\hline
 $N_{\gamma \gamma}$  & 20255$\pm$398 & 1676$\pm$109 & 18572$\pm$370 \\
 $N_{\gamma j}$       & ~~$\!$2575$\pm$516 & ~~$\!$317$\pm$148 & ~~$\!$2217$\pm$459 \\ 
 $N_{jj}$             & 10992$\pm$344 & 854$\pm$96 & 10185$\pm$314 \\
 $N_{\rm ZDY}$        & ~~$\!$198$\pm$14 &   ~~$\!$2.7$\pm$1.7 & ~~$\!$195$\pm$13 \\
 Total                & 34020 & $\!\!$2851 & 31169 \\\hline 
  \end{tabular}
\end{table}

The estimated numbers of DDP events in each bin are corrected for 
the geometric and kinematic acceptance
of the photon, as well as for the photon detection efficiency.
Both acceptance and efficiency are calculated using {\sc sherpa} MC events. 
The acceptance is calculated for the events satisfying
at the particle level \cite{particle}  
$p_T^\gamma>18 ~(17)$ GeV for the leading (next-to-leading) photon, $|\eta^\gamma|<0.9$, and $\Delta \mathcal R_{\gamma \gamma}>0.4$. 
The photon is also required to be isolated by $p_{T}^{\rm iso} =p_{T}^{\rm tot}(0.4)-p_{T}^\gamma < 2.5$ GeV,
where $p_{T}^{\rm tot}(0.4)$ is the scalar sum of the transverse momenta of 
the stable particles within a cone of radius ${\cal R}=0.4$ centered on the photon. 
The acceptance is driven by selection requirements in $\eta_{\rm det}$
(applied to avoid edge effects in the calorimeter regions used for the measurement)
and $\phi_{\rm det}$ (to avoid periodic calorimeter module boundaries)~\cite{d0det},
photon rapidity $\eta^\gamma$ and $p_T$, and bin-to-bin migration effects 
due to the finite energy and angular resolution of the EM calorimeter.
Typically, greater than $80\%$ of events at the reconstruction level remain in the same bin 
as at the particle level.
Choice of an incorrect event vertex leads to a systematic uncertainty 
on the acceptance, typically $\lesssim3\%$ for $\Delta\phi_{\gamma\gamma}\geq\pi/2$ and 
$\lesssim 6\%$ for $\Delta\phi_{\gamma\gamma}<\pi/2$.
The systematic uncertainty is estimated by using DDP events simulated with {\sc sherpa} in which
the event vertex position is randomized according to its distribution in $z$
with respect to the true vertex, and by recalculating 
all relevant variables of the diphoton system.
%
Possible model-dependent effects are corrected by recalculating the acceptance according 
to the difference between the photon $p_T$ spectra in data and {\sc sherpa} MC.
The acceptance grows from 45\% in the low \mass \space region to 80\% in the high mass region.
The systematic uncertainty on the acceptance varies within (4--21)\%. 
In the regions dominated by fragmentation photons, such as low $\Delta\phi_{\gamma\gamma}$ and intermediate \qt,
the acceptance is lower than in the regions dominated by direct production.
The EM clusters reconstructed in the acceptance region are
required to pass the photon identification criteria listed above.
%
%
Small differences between the photon identification
efficiencies in data and MC are corrected by using control samples of electrons from $Z$ boson decays and photons 
from radiative $Z$ boson decays~\cite{nnvalid, d0_dpp}. 
The overall diphoton selection efficiency is typically 
about 50\% with variations of $\pm5\%$.
The relative systematic uncertainty of the diphoton selection efficiency is about 4\%. 

\begin{table}[h]
\begin{center}
\small
\caption{The measured differential cross sections in bins of \mass ~and \qt. The columns
$\delta_{\text {stat}}$, $\delta_{\text {syst}}$, and $\delta_{\text {tot}}$ represent the statistical, systematic, and total uncertainties, respectively.
}
\label{tab:sigmas_gen1}
\begin{tabular}{l@{\hspace{0.6mm}}c@{\hspace{-0.4mm}}c@{\hspace{1.0mm}}c@{\hspace{1.5mm}}c@{\hspace{1.5mm}}c} \hline 
~$M_{\gamma\gamma}$ &~~~$\langle M_{\gamma\gamma}\rangle$~~ & $d\sigma/dM_{\gamma\gamma}$ & $\delta_{\text {stat}}$ &  $\delta_{\text {syst}}$
& ~~$\delta_{\text {tot}}$ \\ 
(GeV) & (GeV)& (pb/GeV) &  (\%) &  (\%) &   (\%) \\\hline
30--40& 37.0 & 1.47$\times10^{-1}$ & 8 & $+15$/$-11$ & $+17$/$-14$  \\
40--50& 44.8 & 3.06$\times10^{-1}$ & 4 & $+14$/$-10$ & $+15$/$-11$  \\
50--60& 54.5 & 1.44$\times10^{-1}$ & 4 & $+11$/$-9$ & $+12$/$-10$  \\
60--70& 64.5 & 7.93$\times10^{-2}$ & 5 & $+11$/$-9$ & $+12$/$-11$  \\
70--80& 74.6 & 4.21$\times10^{-2}$ & 7 & $+14$/$-12$ & $+16$/$-14$  \\
80--90& 84.6 & 2.57$\times10^{-2}$ & 7 & $+13$/$-11$ & $+14$/$-12$  \\
90--100& 94.8 & 1.53$\times10^{-2}$ & 9 & $+14$/$-13$ & $+16$/$-15$  \\
100--125& 110.9 & 7.97$\times10^{-3}$ & 6 & $+12$/$-10$ & $+14$/$-12$  \\
125--150& 136.2 & 2.88$\times10^{-3}$ & 7 & $+15$/$-14$ & $+16$/$-16$  \\
150--200& 170.4 & 1.27$\times10^{-3}$ & 7 & $+15$/$-13$ & $+16$/$-15$  \\
200--350& 249.2 & 2.66$\times10^{-4}$ & 8 & $+15$/$-14$ & $+17$/$-17$  \\
350--500& 403.0 & 2.70$\times10^{-5}$ & 22 & $+47$/$-47$ & $+53$/$-52$  \\\hline \\\hline
~$p_T^{\gamma\gamma}$ &~~~$\langle p_T^{\gamma\gamma}\rangle$~~ & $d\sigma/dp_T^{\gamma\gamma}$ & ~~$\delta_{\text {stat}}$ &  $\delta_{\text {syst}}$
& ~~$\delta_{\text {tot}}$ \\ 
(GeV) & (GeV)& (pb/GeV) &  (\%) &  (\%) &   (\%) \\\hline
0.0--2.5&  1.4 & 6.27$\times10^{-1}$ & 4 & $+9$/$-9$ & $+10$/$-10$ \\
2.5--5.0&  3.6 & 5.38$\times10^{-1}$ & 12 & $+9$/$-9$ & $+15$/$-15$ \\
5.0--7.5&  6.2 & 3.50$\times10^{-1}$ & 14 & $+10$/$-9$ & $+17$/$-17$ \\
7.5--10&  8.8 & 3.47$\times10^{-1}$ & 15 & $+15$/$-10$ & $+22$/$-19$ \\
10--12.5& 11.2 & 2.35$\times10^{-1}$ & 12 & $+12$/$-11$ & $+17$/$-17$ \\
12.5--15& 13.7 & 1.77$\times10^{-1}$ & 16 & $+12$/$-11$ & $+20$/$-20$ \\
15--20& 17.3 & 1.26$\times10^{-1}$ & 10 & $+12$/$-11$ & $+16$/$-15$ \\
20--25& 22.4 & 6.99$\times10^{-2}$ & 8 & $+12$/$-11$ & $+15$/$-14$ \\
25--30& 27.4 & 5.29$\times10^{-2}$ & 10 & $+12$/$-10$ & $+16$/$-15$ \\
30--40& 34.8 & 6.32$\times10^{-2}$ & 8 & $+12$/$-11$ & $+14$/$-14$ \\
40--50& 44.5 & 5.04$\times10^{-2}$ & 9 & $+13$/$-13$ & $+16$/$-16$ \\
50--60& 54.7 & 2.53$\times10^{-2}$ & 13 & $+13$/$-12$ & $+19$/$-19$ \\
60--80& 67.9 & 1.04$\times10^{-2}$ & 12 & $+12$/$-11$ & $+17$/$-17$ \\
80--100& 87.7 & 3.45$\times10^{-3}$ & 17 & $+20$/$-20$ & $+26$/$-26$ \\
100--120& 108.4 & 1.19$\times10^{-3}$ & 19 & $+20$/$-19$ & $+28$/$-28$ \\
120--170& 139.6 & 4.75$\times10^{-4}$ & 20 & $+20$/$-20$ & $+29$/$-28$ \\\hline 
\end{tabular}
\end{center}
\end{table}

\begin{table}[htbp]
\begin{center}
\small
\caption{The measured differential cross sections in bins of \dphi ~and \cost. The columns
$\delta_{\text {stat}}$, $\delta_{\text {syst}}$ and $\delta_{\text {tot}}$ represent the statistical, systematic, and total uncertainties, respectively.
}
\label{tab:sigmas_gen2}
\begin{tabular}{l@{\hspace{-0.6mm}}c@{\hspace{-0.4mm}}c@{\hspace{-1.0mm}}c@{\hspace{1.5mm}}c@{\hspace{1.5mm}}c} \hline 
~~$\Delta\phi_{\gamma\gamma}$ &~~~$\langle \Delta\phi_{\gamma\gamma}\rangle$~~ & $d\sigma/d\Delta\phi_{\gamma\gamma}$ 
& ~~$\delta_{\text {stat}}$ &  $\delta_{\text {syst}}$ & ~~$\delta_{\text {tot}}$ \\ 
 \hspace{0.6mm} (rad) & (rad)& (pb/rad) &  (\%) &  (\%) &   (\%) \\\hline
 0.00--0.31 & 0.17 & 2.28 & 22 & $+12$/$-12$ & $+25$/$-25$  \\
 0.31--0.63 & 0.46 & 1.93 & 16 & $+14$/$-13$ & $+21$/$-21$  \\
 0.63--0.94 & 0.79 & 5.66$\times10^{-1}$ & 12 & $+21$/$-21$ & $+25$/$-24$  \\
 0.94--1.26 & 1.11 & 4.09$\times10^{-1}$ & 13 & $+21$/$-19$ & $+25$/$-23$  \\
 1.26--1.57 & 1.42 & 5.62$\times10^{-1}$ & 20 & $+18$/$-17$ & $+27$/$-26$  \\
 1.57--1.88 & 1.73 & 6.82$\times10^{-1}$ & 11 & $+16$/$-14$ & $+20$/$-18$  \\
 1.88--2.20 & 2.05 & 1.04 & 8 & $+14$/$-13$ & $+17$/$-15$  \\
 2.20--2.51 & 2.37 & 1.65 & 11 & $+14$/$-12$ & $+18$/$-17$  \\
 2.51--2.67 & 2.60 & 3.57 & 13 & $+21$/$-11$ & $+25$/$-17$  \\
 2.67--2.83 & 2.75 & 4.98 & 7 & $+13$/$-11$ & $+14$/$-13$  \\
 2.83--2.98 & 2.91 & 1.08$\times10^{1}$ & 6 & $+13$/$-9$ & $+15$/$-11$  \\
 2.98--3.14 & 3.08 & 2.75$\times10^{1}$ & 3 & $+9$/$-8$ & $+9$/$-9$  \\\hline \\\hline
\cost &~~~$\langle$\cost$\rangle$~~ & $d\sigma/d$\cost
& ~~$\delta_{\text {stat}}$ &  $\delta_{\text {syst}}$ & ~~$\delta_{\text {tot}}$ \\ 
  & & (pb) &  (\%) &  (\%) &   (\%) \\\hline
0.0--0.1& 0.05 & 2.58$\times10^1$ & 6 & $+9$/$-8$ & $+11$/$-10$  \\
0.1--0.2& 0.15 & 2.22$\times10^1$ & 4 & $+9$/$-9$ & $+10$/$-10$  \\
0.2--0.3& 0.25 & 1.91$\times10^1$ & 5 & $+10$/$-9$ & $+11$/$-10$  \\
0.3--0.4& 0.35 & 1.49$\times10^1$ & 5 & $+9$/$-9$ & $+11$/$-10$  \\
0.4--0.5& 0.45 & 9.91 & 7 & $+10$/$-9$ & $+12$/$-12$  \\
0.5--0.6& 0.54 & 5.20 & 9 & $+11$/$-10$ & $+14$/$-14$  \\
0.6--0.7& 0.64 & 1.73 & 12 & $+17$/$-17$ & $+21$/$-21$  \\\hline 
\end{tabular}
\end{center}
\end{table}

\begin{table}[htbp]
\begin{center}
\small
\caption{The measured differential cross sections in bins of \mass, \qt, and $|\cos \theta^{*}|$ for \dphi$<\pi/2$. The columns
$\delta_{\text {stat}}$, $\delta_{\text {syst}}$ and $\delta_{\text {tot}}$ represent the statistical, 
systematic, and total uncertainties, respectively.
}
\label{tab:sigmas_low}
\vspace*{-2mm}
\begin{tabular}{l@{\hspace{0.6mm}}c@{\hspace{-0.4mm}}c@{\hspace{1.0mm}}c@{\hspace{1.5mm}}c@{\hspace{1.5mm}}c} \hline 
~$M_{\gamma\gamma}$ &~~~$\langle M_{\gamma\gamma}\rangle$~~ & $d\sigma/dM_{\gamma\gamma}$ & $\delta_{\text {stat}}$ &  $\delta_{\text {syst}}$
& ~~$\delta_{\text {tot}}$ \\ 
(GeV) & (GeV)& (pb/GeV) &  (\%) &  (\%) &   (\%) \\\hline
30--40& 34.3 & 1.64$\times10^{-2}$ & 14 & $+14$/$-14$ & $+20$/$-20$  \\
40--50& 44.8 & 8.92$\times10^{-3}$ & 28 & $+15$/$-14$ & $+31$/$-31$  \\
50--60& 54.6 & 2.25$\times10^{-3}$ & 23 & $+23$/$-23$ & $+33$/$-33$  \\
60--70& 64.6 & 1.22$\times10^{-3}$ & 41 & $+25$/$-27$ & $+48$/$-49$  \\
70--90& 78.7 & 5.60$\times10^{-4}$ & 30 & $+14$/$-14$ & $+33$/$-33$  \\
90--200& 111.4 & 5.44$\times10^{-5}$ & 41 & $+30$/$-30$ & $+51$/$-51$  \\\hline \\\hline 
~$p_T^{\gamma\gamma}$ &~~~$\langle p_T^{\gamma\gamma}\rangle$~~ & $d\sigma/dp_T^{\gamma\gamma}$ & ~~$\delta_{\text {stat}}$ &  $\delta_{\text {syst}}$
& ~~$\delta_{\text {tot}}$ \\ 
(GeV) & (GeV)& (pb/GeV) &  (\%) &  (\%) &   (\%) \\\hline
25--30& 28.3 & 5.89$\times10^{-3}$ & 44 & $+30$/$-28$ & $+54$/$-53$ \\
30--40& 35.8 & 3.56$\times10^{-2}$ & 23 & $+14$/$-14$ & $+27$/$-27$ \\
40--50& 44.5 & 4.39$\times10^{-2}$ & 15 & $+17$/$-17$ & $+22$/$-22$ \\
50--60& 54.8 & 1.72$\times10^{-2}$ & 18 & $+14$/$-14$ & $+23$/$-23$ \\
60--80& 67.8 & 7.74$\times10^{-3}$ & 17 & $+12$/$-12$ & $+21$/$-21$ \\
80--100& 87.5 & 2.70$\times10^{-3}$ & 22 & $+17$/$-17$ & $+28$/$-28$ \\
100--120& 108.3 & 7.07$\times10^{-4}$ & 17 & $+22$/$-22$ & $+28$/$-28$ \\
120--170& 140.5 & 3.84$\times10^{-4}$ & 25 & $+26$/$-26$ & $+36$/$-36$ \\\hline \\\hline
\cost &~~~$\langle$\cost$\rangle$~~ & $d\sigma/d$\cost
& ~~$\delta_{\text {stat}}$ &  $\delta_{\text {syst}}$ & ~~$\delta_{\text {tot}}$ \\ 
  & & (pb) &  (\%) &  (\%) &   (\%) \\\hline
0.0-0.1& 0.05 & 2.64 & 23 & $+23$/$-23$ & $+33$/$-33$  \\
0.1-0.2& 0.15 & 3.22 & 13 & $+18$/$-18$ & $+23$/$-23$  \\
0.2-0.3& 0.25 & 3.71 & 15 & $+14$/$-14$ & $+21$/$-21$  \\
0.3-0.4& 0.34 & 2.17 & 17 & $+12$/$-12$ & $+21$/$-21$  \\
0.4-0.5& 0.45 & 1.09 & 17 & $+15$/$-15$ & $+23$/$-23$  \\
0.5-0.6& 0.54 & 6.12$\times10^{-1}$ & 39 & $+23$/$-23$ & $+45$/$-45$  \\
0.6-0.7& 0.63 & 3.33$\times10^{-1}$ & 39 & $+27$/$-27$ & $+48$/$-48$  \\\hline 
\end{tabular}
\end{center}
\end{table}

\begin{table}[htbp]
\begin{center}
\small
\caption{The measured differential cross sections in bins of \mass, \qt, and $|\cos \theta^{*}|$ for \dphi$\geq\pi/2$. The columns
$\delta_{\text {stat}}$, $\delta_{\text {syst}}$ and $\delta_{\text {tot}}$ represent the statistical, systematic and total uncertainties, respectively.
}
\label{tab:sigmas_hi}
\vspace*{-2mm}
\begin{tabular}{l@{\hspace{0.6mm}}c@{\hspace{-0.4mm}}c@{\hspace{1.0mm}}c@{\hspace{1.5mm}}c@{\hspace{1.5mm}}c} \hline 
~$M_{\gamma\gamma}$ &~~~$\langle M_{\gamma\gamma}\rangle$~~ & $d\sigma/dM_{\gamma\gamma}$ & $\delta_{\text {stat}}$ &  $\delta_{\text {syst}}$
& ~~$\delta_{\text {tot}}$ \\ 
(GeV) & (GeV)& (pb/GeV) &  (\%) &  (\%) &   (\%) \\\hline
30--40& 37.5 & 1.31$\times10^{-1}$ & 9 & $+11$/$-9$ & $+14$/$-13$  \\
40--50& 44.8 & 2.96$\times10^{-1}$ & 5 & $+9$/$-8$ & $+10$/$-9$  \\
50--60& 54.5 & 1.43$\times10^{-1}$ & 4 & $+8$/$-8$ & $+10$/$-10$  \\
60--70& 64.5 & 8.06$\times10^{-2}$ & 5 & $+9$/$-9$ & $+10$/$-10$  \\
70--80& 74.6 & 4.10$\times10^{-2}$ & 8 & $+9$/$-9$ & $+12$/$-12$  \\
80--90& 84.6 & 2.53$\times10^{-2}$ & 7 & $+12$/$-12$ & $+13$/$-13$  \\
90--100& 94.7 & 1.53$\times10^{-2}$ & 9 & $+12$/$-12$ & $+14$/$-14$  \\
100--125& 110.9 & 7.97$\times10^{-3}$ & 7 & $+10$/$-10$ & $+12$/$-13$  \\
125--150& 136.2 & 2.82$\times10^{-3}$ & 8 & $+15$/$-15$ & $+17$/$-17$  \\
150--200& 170.4 & 1.26$\times10^{-3}$ & 7 & $+13$/$-13$ & $+15$/$-15$  \\
200--350& 249.2 & 2.65$\times10^{-4}$ & 9 & $+15$/$-15$ & $+17$/$-17$  \\
350--500& 403.0 & 2.70$\times10^{-5}$ & 28 & $+47$/$-47$ & $+55$/$-55$  \\\hline \\\hline
~$p_T^{\gamma\gamma}$ &~~~$\langle p_T^{\gamma\gamma}\rangle$~~ & $d\sigma/dp_T^{\gamma\gamma}$ & ~~$\delta_{\text {stat}}$ &  $\delta_{\text {syst}}$
& ~~$\delta_{\text {tot}}$ \\ 
(GeV) & (GeV)& (pb/GeV) &  (\%) &  (\%) &   (\%) \\\hline
0.0--2.5&  1.4 & 6.34$\times10^{-1}$ & 5 & $+9$/$-9$ & $+10$/$-10$ \\
2.5--5.0&  3.6 & 5.42$\times10^{-1}$ & 16 & $+8$/$-8$ & $+18$/$-18$ \\
5.0--7.5&  6.2 & 3.52$\times10^{-1}$ & 20 & $+9$/$-9$ & $+22$/$-22$ \\
7.5--10&  8.8 & 3.48$\times10^{-1}$ & 14 & $+12$/$-9$ & $+19$/$-17$ \\
10--12.5& 11.2 & 2.35$\times10^{-1}$ & 17 & $+11$/$-11$ & $+20$/$-20$ \\
12.5--15& 13.7 & 1.77$\times10^{-1}$ & 16 & $+11$/$-11$ & $+20$/$-20$ \\
15--20& 17.3 & 1.26$\times10^{-1}$ & 11 & $+10$/$-10$ & $+15$/$-15$ \\
20--25& 22.4 & 6.96$\times10^{-2}$ & 9 & $+10$/$-10$ & $+14$/$-14$ \\
25--30& 27.3 & 4.82$\times10^{-2}$ & 11 & $+12$/$-12$ & $+16$/$-17$ \\
30--40& 34.2 & 3.03$\times10^{-2}$ & 8 & $+10$/$-10$ & $+13$/$-13$ \\
40--50& 44.4 & 1.21$\times10^{-2}$ & 10 & $+9$/$-9$ & $+14$/$-14$ \\
50--60& 54.5 & 5.93$\times10^{-3}$ & 14 & $+20$/$-20$ & $+24$/$-24$ \\
60--80& 68.2 & 2.18$\times10^{-3}$ & 14 & $+16$/$-16$ & $+21$/$-21$ \\
80--100& 88.3 & 5.83$\times10^{-4}$ & 25 & $+29$/$-29$ & $+38$/$-38$ \\
100--120& 108.5 & 4.91$\times10^{-4}$ & 27 & $+24$/$-24$ & $+36$/$-36$ \\
120--170& 137.4 & 1.13$\times10^{-4}$ & 30 & $+31$/$-31$ & $+43$/$-43$ \\\hline\\\hline
\cost &~~~$\langle$\cost$\rangle$~~ & $d\sigma/d$\cost
& ~~$\delta_{\text {stat}}$ &  $\delta_{\text {syst}}$ & ~~$\delta_{\text {tot}}$ \\ 
  & & (pb) &  (\%) &  (\%) &   (\%) \\\hline
0.0-0.1& 0.05 & 2.24$\times10^{1}$ & 6 & $+8$/$-8$ & $+10$/$-10$  \\
0.1-0.2& 0.15 & 1.86$\times10^{1}$ & 5 & $+8$/$-8$ & $+9$/$-9$  \\
0.2-0.3& 0.25 & 1.55$\times10^{1}$ & 6 & $+9$/$-8$ & $+11$/$-11$  \\
0.3-0.4& 0.35 & 1.24$\times10^{1}$ & 6 & $+8$/$-8$ & $+10$/$-10$  \\
0.4-0.5& 0.45 & 8.38 & 8 & $+9$/$-9$ & $+12$/$-12$  \\
0.5-0.6& 0.54 & 4.43 & 10 & $+11$/$-10$ & $+15$/$-14$  \\
0.6-0.7& 0.64 & 1.04 & 17 & $+22$/$-22$ & $+28$/$-28$  \\\hline 
\end{tabular}
\end{center}
\end{table}

The differential cross sections \dmass, \dqt, \ddphi, and \dcost \space 
are calculated from the number of data events after the subtraction of background contributions
divided by the event selection efficiencies, acceptance, integrated luminosity, and the bin width.

The measured differential cross sections for all considered kinematic regions are 
presented in Tables \ref{tab:sigmas_gen1}--\ref{tab:sigmas_hi}.
The average value of each variable in a bin was estimated using {\sc sherpa} MC events.
The statistical uncertainty $\delta_{\rm stat}$ 
is caused by finite MC statistics used for the efficiency and acceptance calculations
and by the statistical uncertainty 
in data, taking into account statistical correlations with adjacent bins.
The latter are estimated using an inverted smearing matrix, following a procedure described in Ref.~\cite{GamJet_MPI}.
The smearing matrix represents the detector resolution function and relates each bin
at the particle level to the bins at the reconstruction level.
It is constructed for each variable using the DDP MC events simulated with {\sc sherpa}. 


\begin{figure*}[htb]
 \centering
  \includegraphics[scale=0.33]{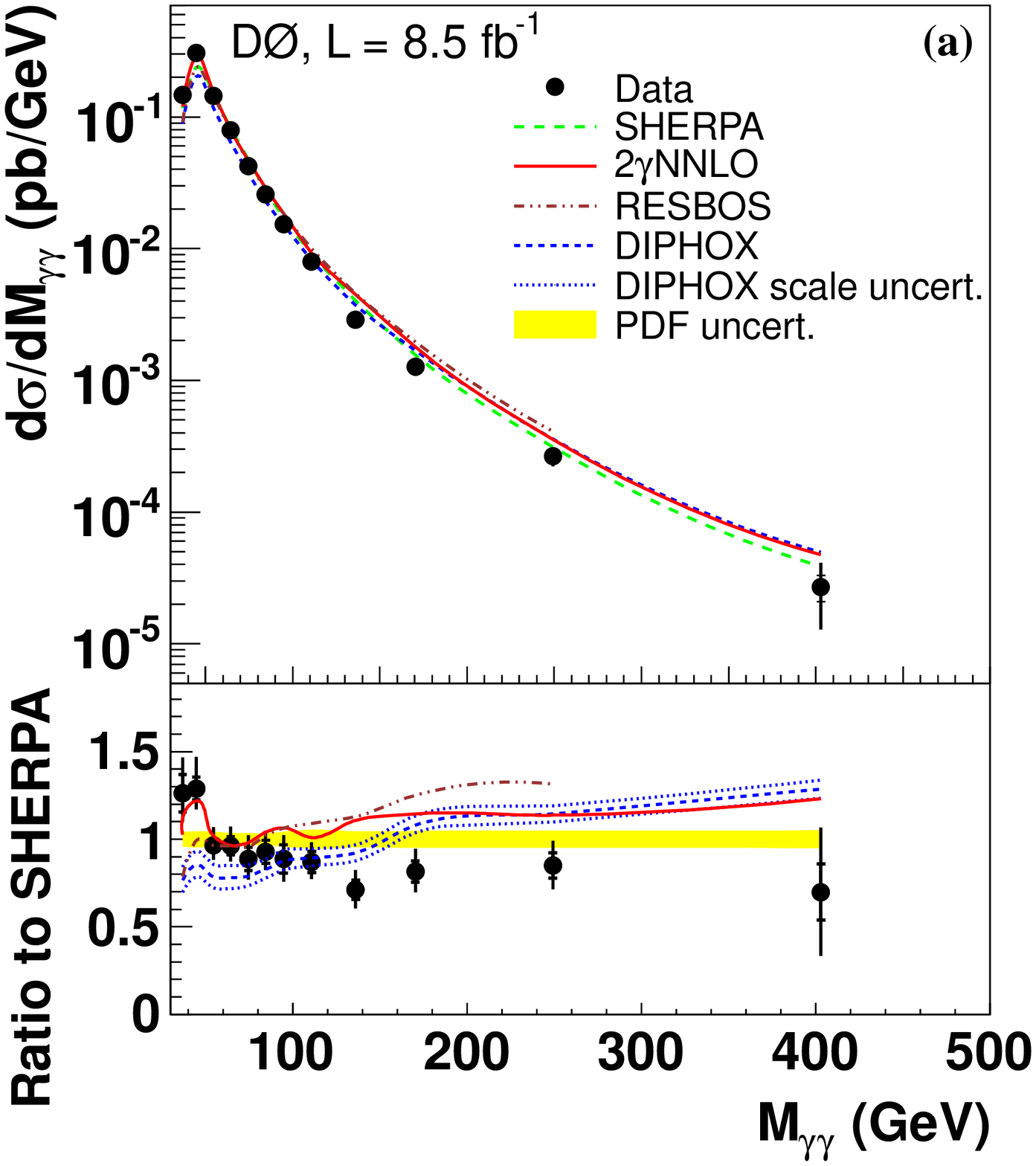}
  \includegraphics[scale=0.33]{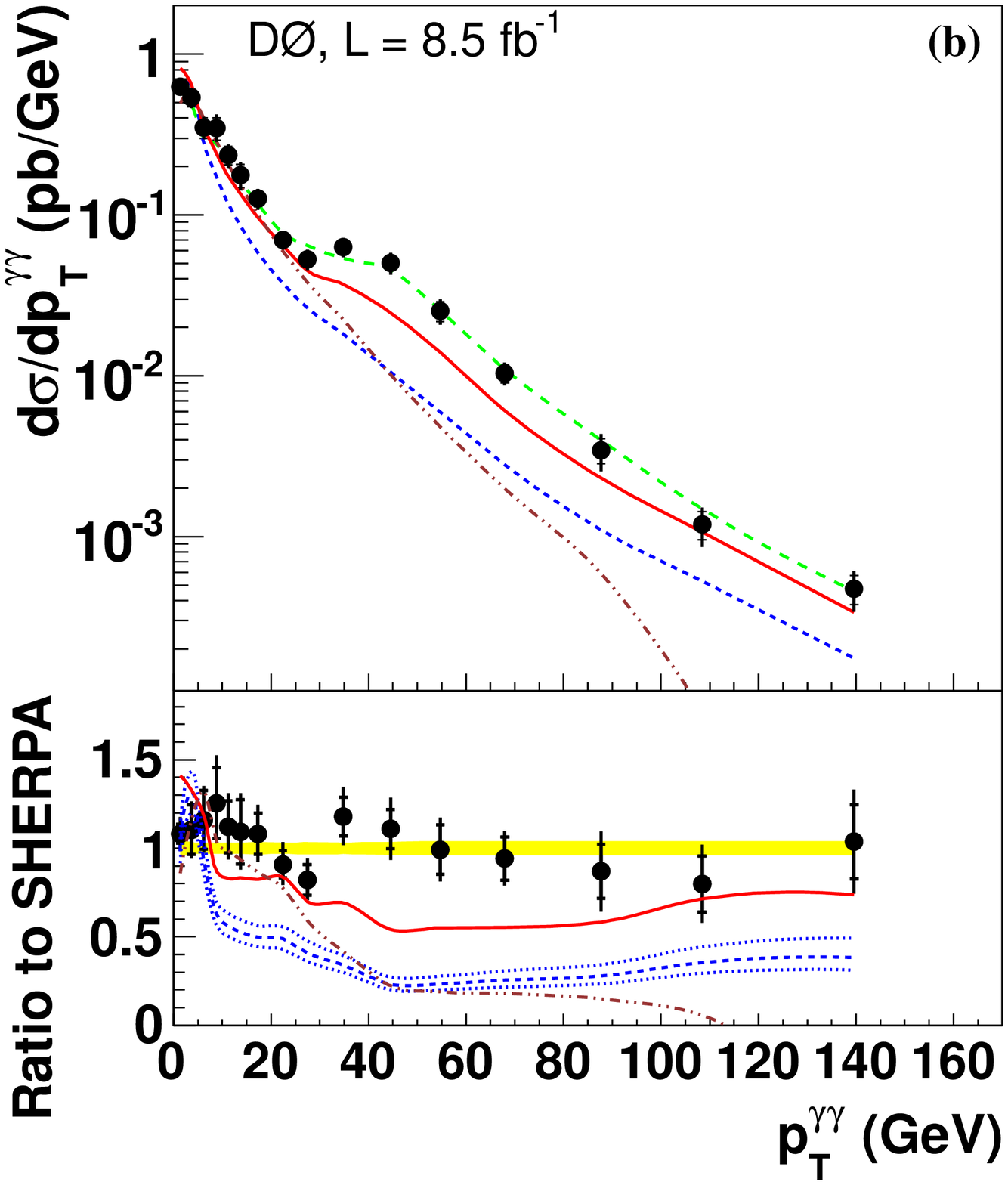}
  \includegraphics[scale=0.33]{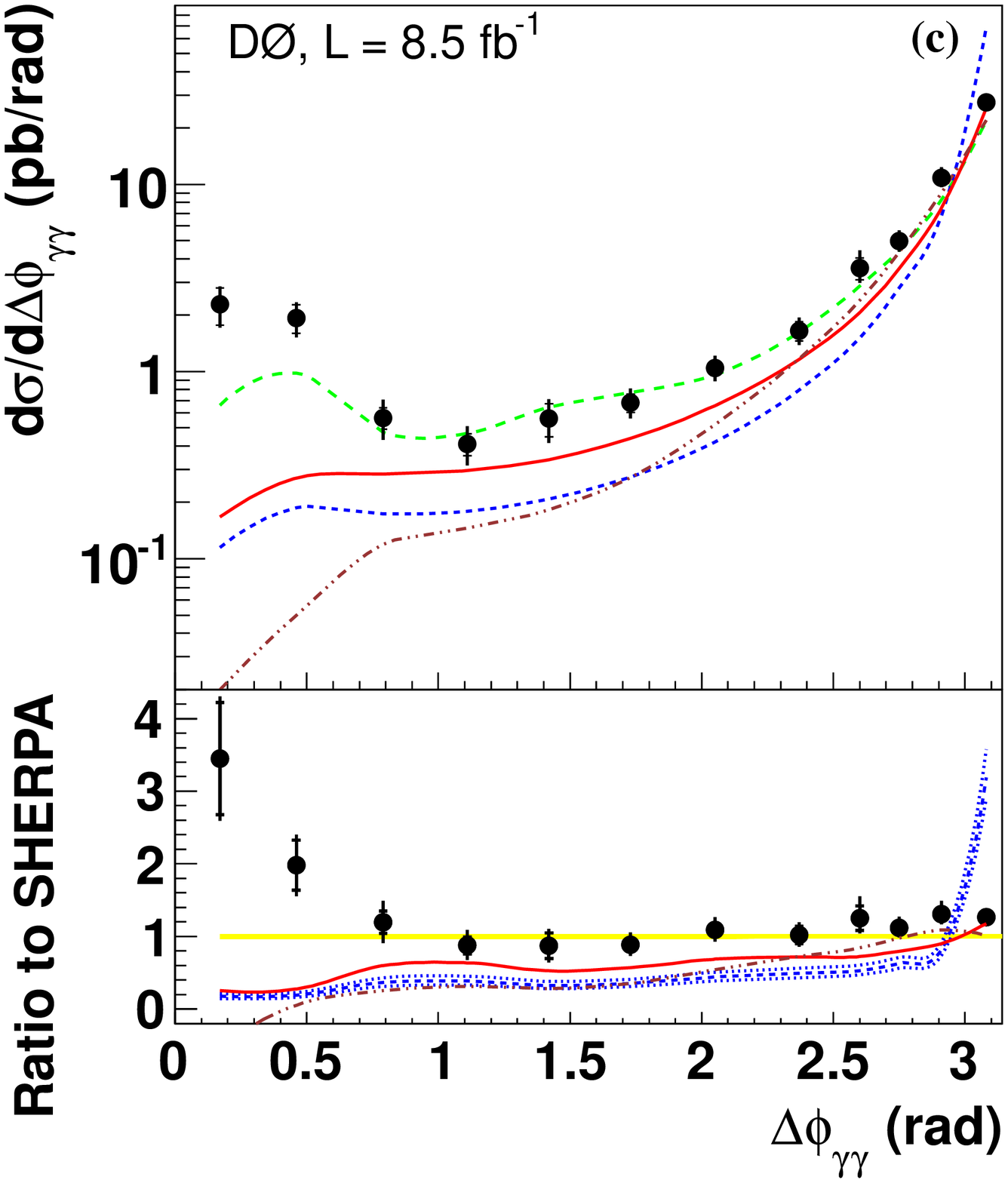}
  \includegraphics[scale=0.33]{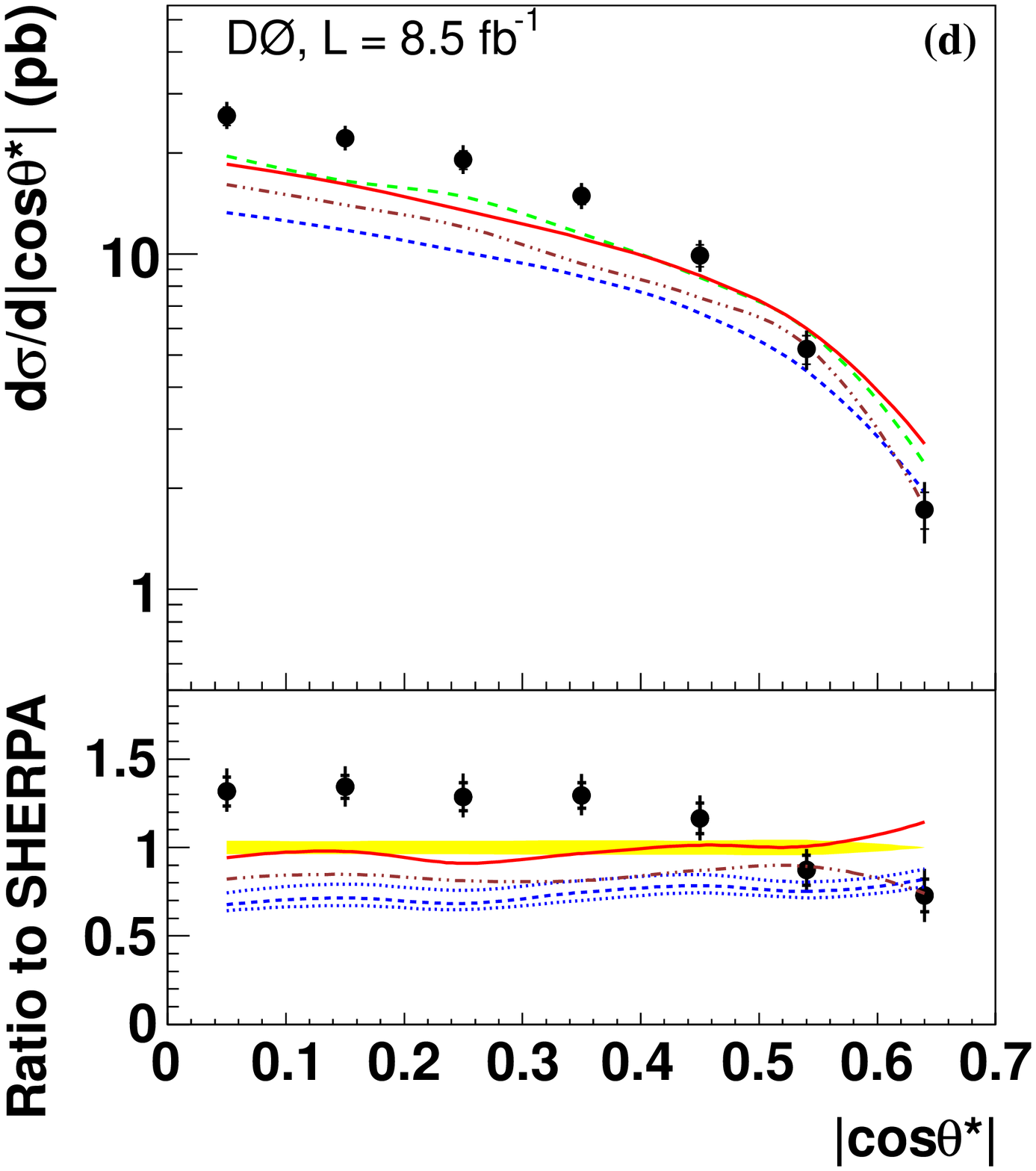}
\vspace*{-5mm}
  \caption{(Color online) The differential cross section
    as a function of (a) $M_{\gamma\gamma}$, (b) $p_{T}^{\gamma\gamma}$, (c) \dphi, and (d) \cost 
   ~for the full \dphi \space region from data (black points) and theory predictions (curves) are shown in the upper plots. 
   The lower plots show the ratio of data and {\sc diphox, resbos}, and $2\gamma${\sc nnlo} 
    predictions to the {\sc sherpa} predictions.
   The inner line for the error bars in data points shows the statistical uncertainty, while the outer line shows 
   the total (statistical and systematic added in quadrature) uncertainty
after subtracting the 7.4\% normalization uncertainty. 
  }
  \label{fig:fulldphi}
\end{figure*}

\begin{figure*}[htb]
 \centering
\hspace*{-4mm}
 \includegraphics[scale=0.31]{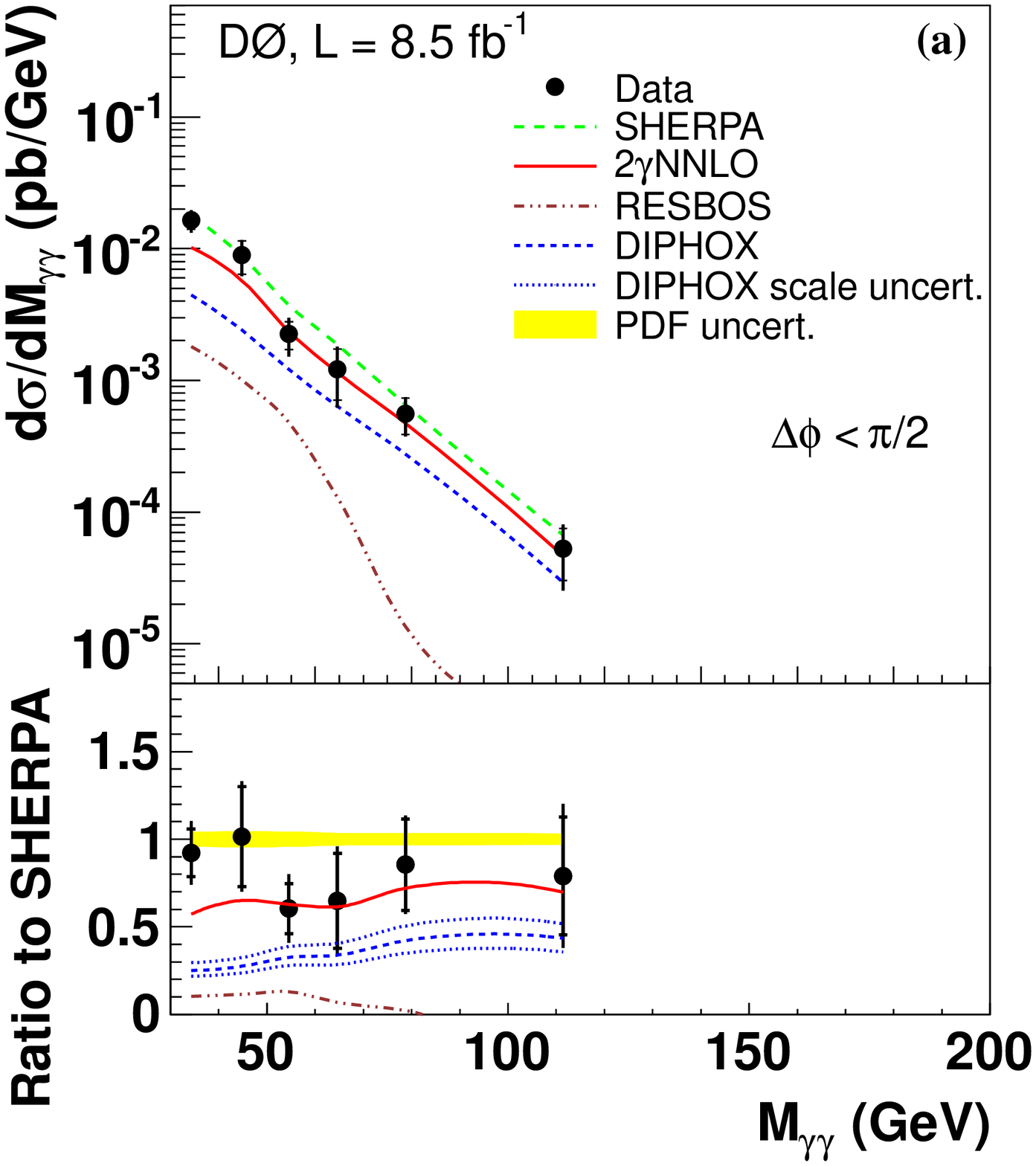}
\hspace*{-5mm}
 \includegraphics[scale=0.31]{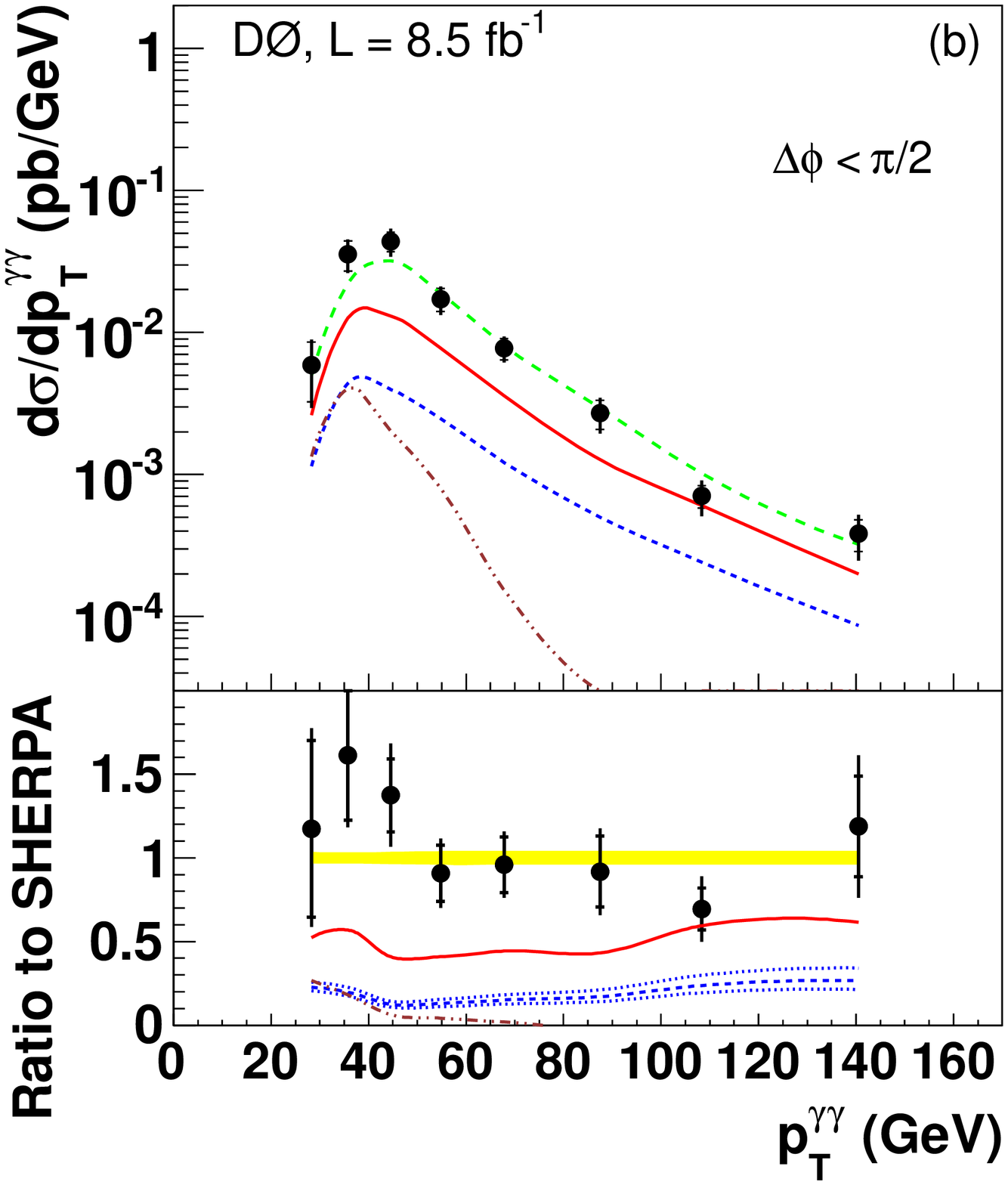}
\hspace*{-5mm}
 \includegraphics[scale=0.31]{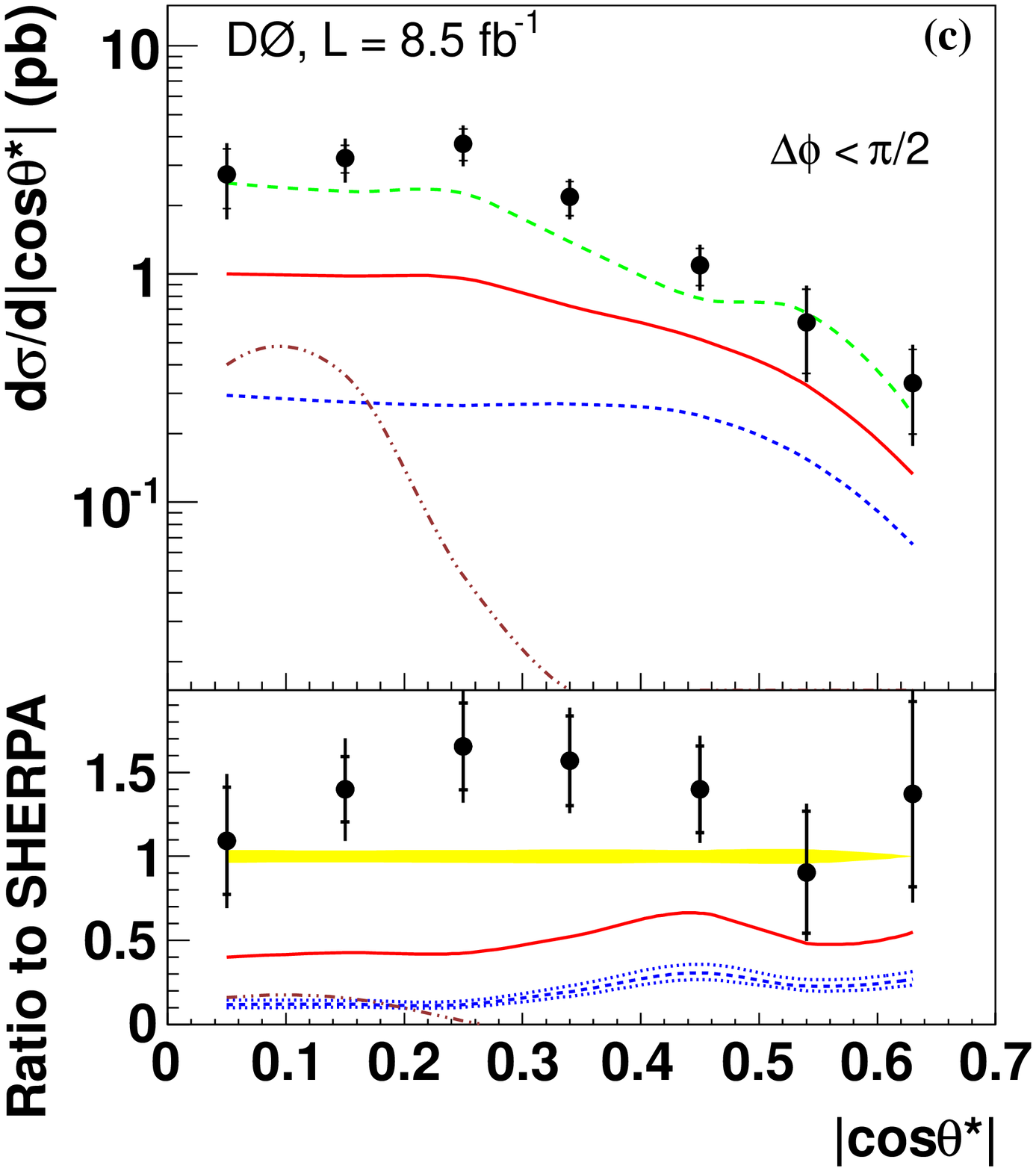}
\vspace*{-5mm}
  \caption{(Color online) The differential cross section
    as a function of (a) $M_{\gamma\gamma}$, (b) $p_{T}^{\gamma\gamma}$, and (c) \cost 
   ~for the \dphi$<\pi/2$ region. The notations for points, lines and shaded regions are the same as in Fig.~\ref{fig:fulldphi}. }
  \label{fig:lowdphi}
\end{figure*}

\begin{figure*}[htbp]
 \centering
\hspace*{-4mm}
 \includegraphics[scale=0.31]{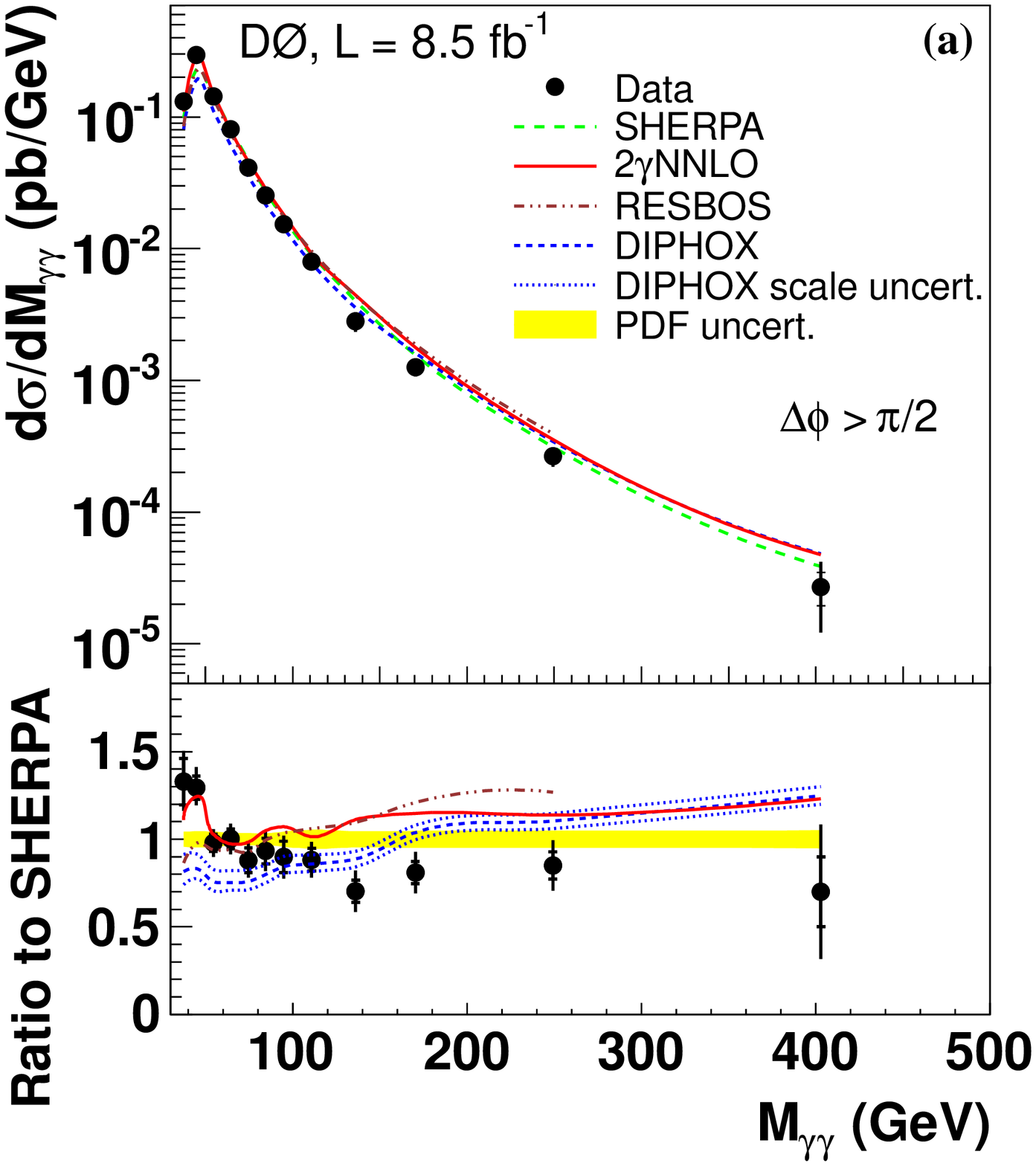}
\hspace*{-5mm}
 \includegraphics[scale=0.31]{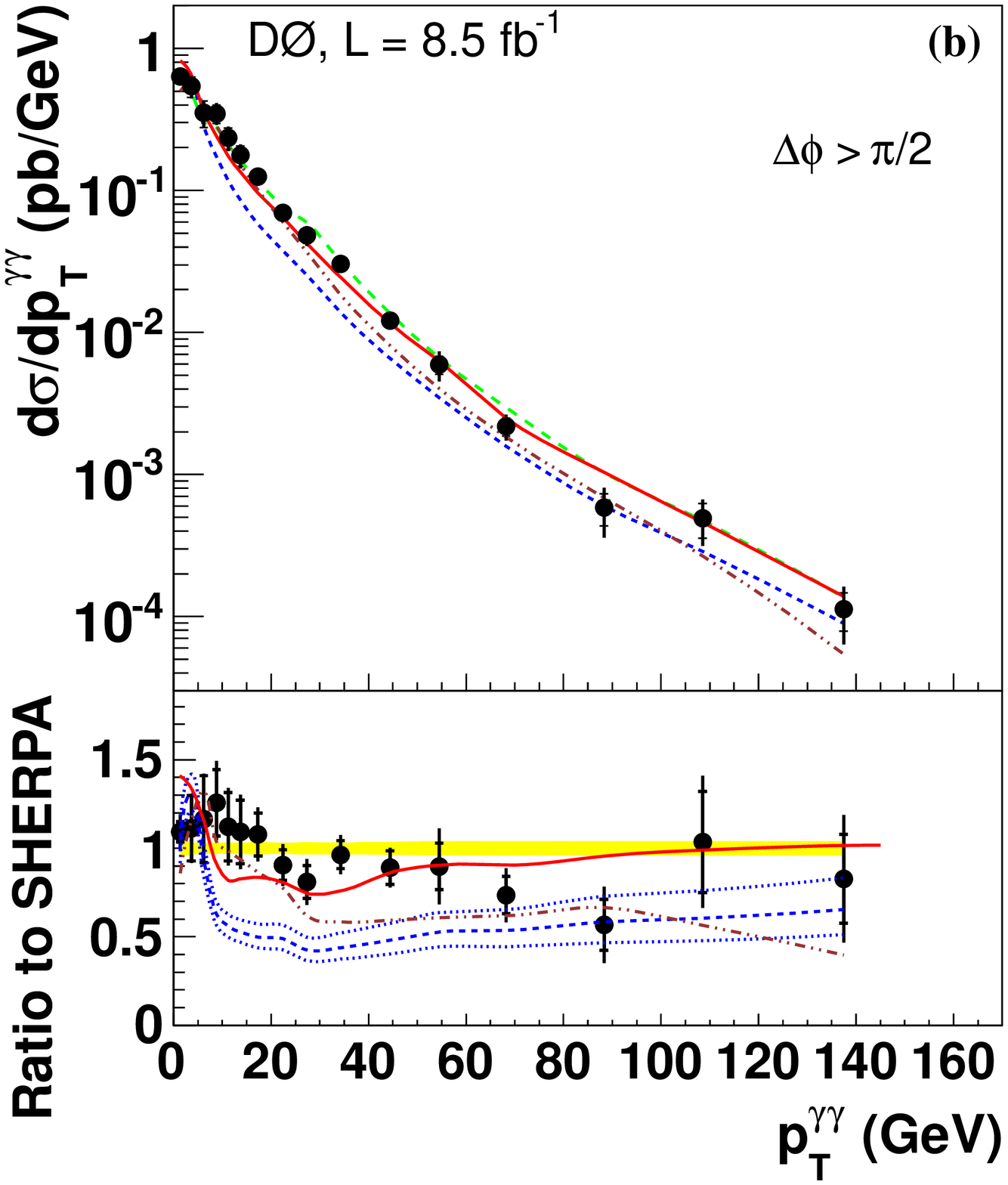}
\hspace*{-5mm}
 \includegraphics[scale=0.31]{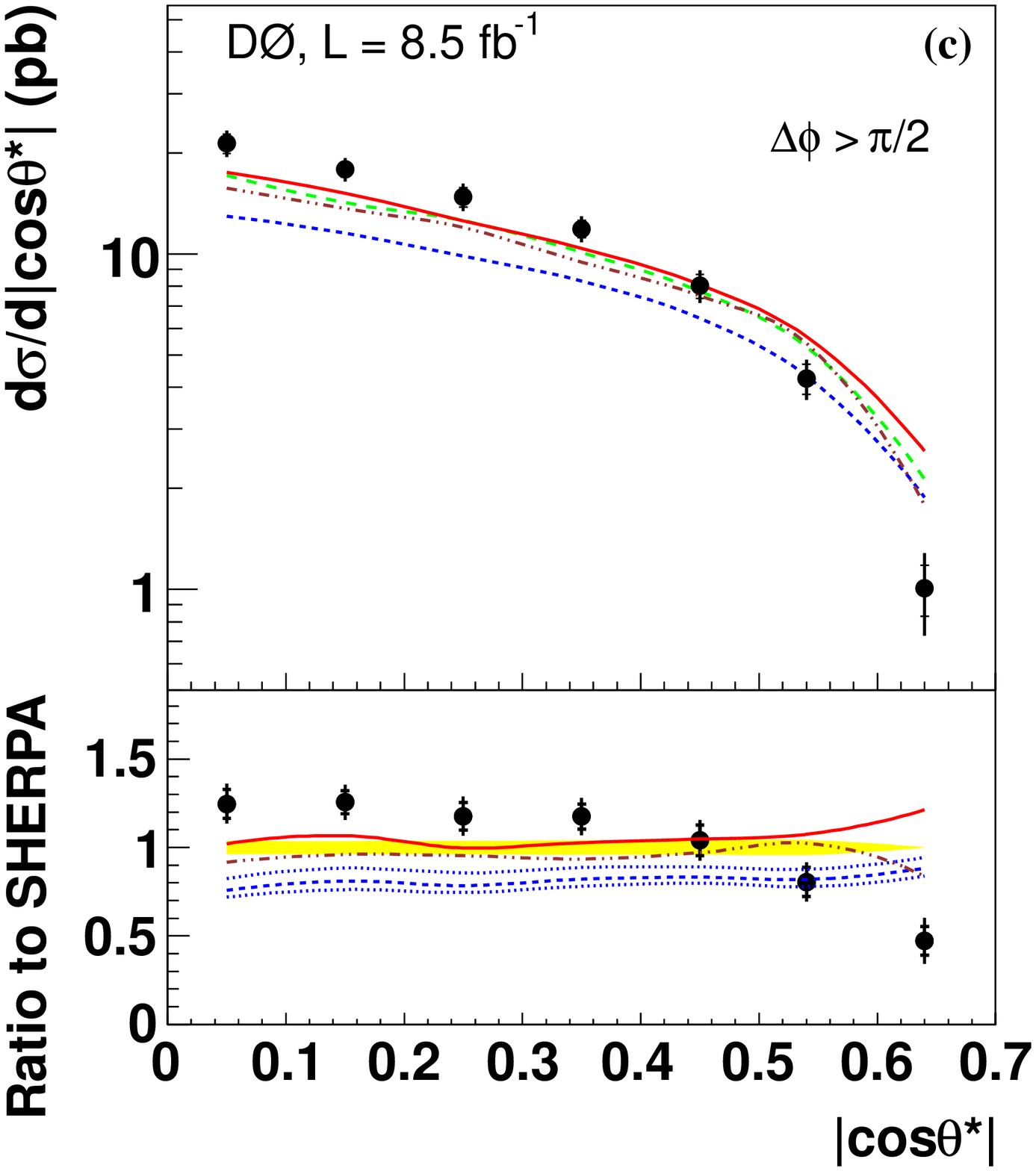}
\vspace*{-5mm}
  \caption{(Color online) The differential cross section
    as a function of (a) $M_{\gamma\gamma}$, (b) $p_{T}^{\gamma\gamma}$, and (c) \cost 
   ~for the \dphi$\geq\pi/2$ region. The notations for points, lines and shaded regions are the same as in Fig.~\ref{fig:fulldphi}. }
  \label{fig:hidphi}
\end{figure*}

Figures~\ref{fig:fulldphi}--\ref{fig:hidphi} show a comparison of the measured differential cross sections to the theoretical predictions 
from  {\sc diphox},  {\sc resbos}, $2\gamma${\sc nnlo}, and {\sc sherpa}.
The {\sc resbos} predictions are valid only for the phase space limited by $9<M_{\gamma\gamma}<350$ GeV.
We take this into account in our calculations
and compare {\sc resbos} predictions to $M_{\gamma\gamma}$ measurements
up to $\approx\!250$ GeV (see Table II), the last mass value below 350 GeV where the cross section is measured.
Systematic uncertainties across the bins in the measured cross sections are largely 
($>\!90\%$) correlated. 
A common normalization uncertainty of 7.4\% resulting from luminosity and diphoton selection efficiency is not shown in the plots. 
The predictions from {\sc sherpa}, {\sc diphox} and  {\sc resbos}
are computed using the {\sc cteq}6.6M NLO PDFs~\cite{cteq}, and from $2\gamma${\sc nnlo} using {\sc mstw2008} NNLO PDFs~\cite{mstw}.
The PDF uncertainty is estimated using {\sc diphox} and the 44 eigenvectors provided with the {\sc cteq}6.6M PDF set. 
They are found to be within (3--7)\%.
The renormalization $\mu_R$, factorization $\mu_F$, and fragmentation $\mu_f$ scales 
are set to $\mu_R=\mu_F=\mu_f=\mbox{\mass}$. 
The uncertainty due to the scale choice is estimated using {\sc diphox} via a simultaneous variation 
by a factor of two of all scales relative to the default values and found to be about 
10\% for $d\sigma/d$\mass \space and $d\sigma/d$\cost, 
and a maximum of (20--28)\% for $d\sigma/d$\qt  \space at high \qt and for $d\sigma/d$\dphi
\space at low \dphi. 
 All theoretical predictions are obtained using diphoton event selection criteria
equivalent to those applied in the experimental analysis (as are those used for the acceptance calculation). 
In particular, the photon is required to be isolated by $p_{T}^{\rm iso} < 2.5$ GeV.
For {\sc diphox},  {\sc resbos}, and $2\gamma${\sc nnlo}, $p_{T}^{\rm tot}$ is computed at the parton level.
The cross sections from {\sc diphox},  {\sc resbos} and $2\gamma${\sc nnlo} are corrected 
for effects stemming from multiple parton interactions and hadronization, 
while for {\sc sherpa} 
these effects are handled within the software package.
These corrections are estimated using 
diphoton events simulated by {\sc pythia} with Tunes A and S0~\cite{pythia}. 
The corrections vary within (4--6)\% as a function of the measured kinematic variables and are consistent for both tunes within 1\%. 

Tables \ref{tab:sigmas_gen1}--\ref{tab:sigmas_hi} show that the cross sections in the \dphi$\geq\pi/2$ region
constitute, on average, about (85--90)\% of the cross sections for the full \dphi ~range. From the sub-tables for the \qt ~variable,
we observe that at \qt$\lesssim 25$ GeV, the cross sections are fully dominated by the \dphi$\geq\pi/2$ region,
while starting from \qt$\gtrsim 30$ GeV, they are significantly dominated (by a factor of 2--4) by the  \dphi$<\pi/2$ region.
The shoulder-like structure observed in the \qt ~distribution around $30-40$ GeV
should be mainly caused by the fragmentation photons coming from the \dphi$<\pi/2$ region, 
and partially by higher-order (NLO and beyond) corrections \cite{diphox_tev}.

In general, none of the theoretical models considered here provides a consistent description of the experimental results 
in all kinematic regions. 
The {\sc sherpa} predictions are able to describe most of the phase space relatively well
except for the low DDP mass region, very low \dphi, and with some tension in the \cost ~spectrum.
A noticeable discrepancy between {\sc resbos} and {\sc diphox}
in some regions of the phase space is due to the absence of all-order soft-gluon resummation 
($p_{T}^{\gamma\gamma}$ close to zero and \dphi ~close to $\pi$) and the fact that the
$gg\to\gamma\gamma$ contribution is calculated only at LO in {\sc diphox} (small $M_{\gamma\gamma}$).
However, {\sc resbos} fails to describe $M_{\gamma\gamma}$, $p_{T}^{\gamma\gamma}$, and \cost  ~spectra
in the \dphi$<\pi/2$ region, where the contributions from the fragmentation diagrams and higher-order corrections are important.
The processes with a parton-to-diphoton fragmentation taking place
at low masses ($M_{\gamma\gamma}<p_{T}^{\gamma\gamma}$) are not included yet in any existing calculation~\cite{resbos}.
The regions of phase space with a significant contribution from fragmentation photons (very low \dphi)
require extensive tuning of all of the considered event generators.

In summary, we have presented 
measurements of differential cross sections of photon pair production 
in $p \bar p$ collisions at $\sqrt{s}=1.96$ TeV as functions of \mass, \qt, \dphi, and \cost ~for photons with $p_T>18(17)$ GeV and $|\eta|<0.9$ in the full \dphi \space range and for \dphi$<\pi/2$, \dphi$\geq \pi/2$ separately.
The cross sections are compared to the predictions made by the {\sc diphox}, {\sc resbos}, $2\gamma${\sc nnlo}
and {\sc sherpa} MC generators. 
Overall, {\sc sherpa} provides the best description of the measured cross sections.
The experimental results show discrepancies with all theoretical predictions in the regions of 
small \dphi ~and small diphoton mass for \dphi$\geq\pi/2$, with minor differences in the shapes of the \cost ~distribution. 
The results are important for understanding of DDP production and tuning of modern generators to study 
SM phenomena and search for beyond the SM processes.

%
We thank the staffs at Fermilab and collaborating institutions,
and acknowledge support from the
DOE and NSF (USA);
CEA and CNRS/IN2P3 (France);
MON, NRC KI and RFBR (Russia);
CNPq, FAPERJ, FAPESP and FUNDUNESP (Brazil);
DAE and DST (India);
Colciencias (Colombia);
CONACyT (Mexico);
NRF (Korea);
FOM (The Netherlands);
STFC and the Royal Society (United Kingdom);
MSMT and GACR (Czech Republic);
BMBF and DFG (Germany);
SFI (Ireland);
The Swedish Research Council (Sweden);
and
CAS and CNSF (China).
%


\begin{thebibliography}{99}
\bibitem{d0_coordinate}
The polar angle $\theta$ is defined
with respect to the positive $z$ axis, which is along the proton beam direction,
while the azimuthal angle $\phi$ is defined with respect to the $x$ axis.
Pseudo-rapidity is defined as $\eta = -\ln[\tan(\theta/2)]$.
Also, $|\eta_{\rm det}|$ and $|\phi_{\rm det}|$ are the pseudorapidity and the azimuthal angle 
defined with respect to the geometric center of the D0 detector.
\end{thebibliography}
\end{document}